\documentclass{cimento}

\usepackage{graphicx}  

\title{Compressed sensing and Sequential Monte Carlo for solar hard X-ray imaging}
\author{A. M.~Massone\from{ins:x}\from{ins:y},
F.~Sciacchitano\from{ins:x},
M.~Piana\from{ins:x}\from{ins:y} \atque
A.~Sorrentino\from{ins:x}\from{ins:y}
}
\instlist{\inst{ins:x} Dipartimento di Matematica, Universit\`a di Genova, Genova, Italy
  \inst{ins:y} CNR - SPIN, Genova, Italy}

\begin{document}

\maketitle

\begin{abstract}
We describe two inversion methods for the reconstruction of hard X-ray solar images. The methods are tested against experimental visibilities recorded by the {\em{Reuven Ramaty High Energy Solar Spectroscopic Imager (RHESSI)}} and synthetic visibilities based on the design of the {\em{Spectrometer/Telescope for Imaging X-rays (STIX)}}. 
\end{abstract}

\section{Introduction}
The NASA {\em{Reuven Ramaty High Energy Solar Spectroscopic Imager (RHESSI)}} \cite{ref:lietal02} and the ESA {\em{Spectrometer/Telescope for Imaging X-rays (STIX)}} \cite{ref:beetal12} are two space telescopes for imaging hard X-rays that rely on rather similar imaging technologies. {\em{RHESSI}} has been decommissioned on August 16 2018 after more than $16$ years of successful operations, while {\em{STIX}} is going to fly in the next two years. Both hardwares allow the modulation of the X-ray flux coming from the Sun, providing as a result sparse samples of its Fourier transform, named visibilities, picked up at specific $(u,v)$ points of the Fourier plane. Therefore, for both {\em{RHESSI}} and {\em{STIX}}, image reconstruction is needed to determine the actual spatial photon flux distribution from the few Fourier components acquired by the hard X-ray collimators \cite{ref:asetal02} \cite{ref:beetal13} \cite{ref:depe09} \cite{ref:feetal17} \cite{ref:maetal09} \cite{ref:meetal96}. In Section 2 of this paper we briefly overview a reconstruction method based on compressed sensing \cite{ref:duetal18}. In Section 3 we provide more insights on a Monte Carlo method for the Bayesian estimation of several imaging parameters \cite{ref:scetal19} \cite{ref:scetal18}. Our conclusions are offered in Section 4.

\section{Compressed sensing for hard X-ray image reconstruction}
\begin{figure}
\begin{center}
\includegraphics[width=12.cm]{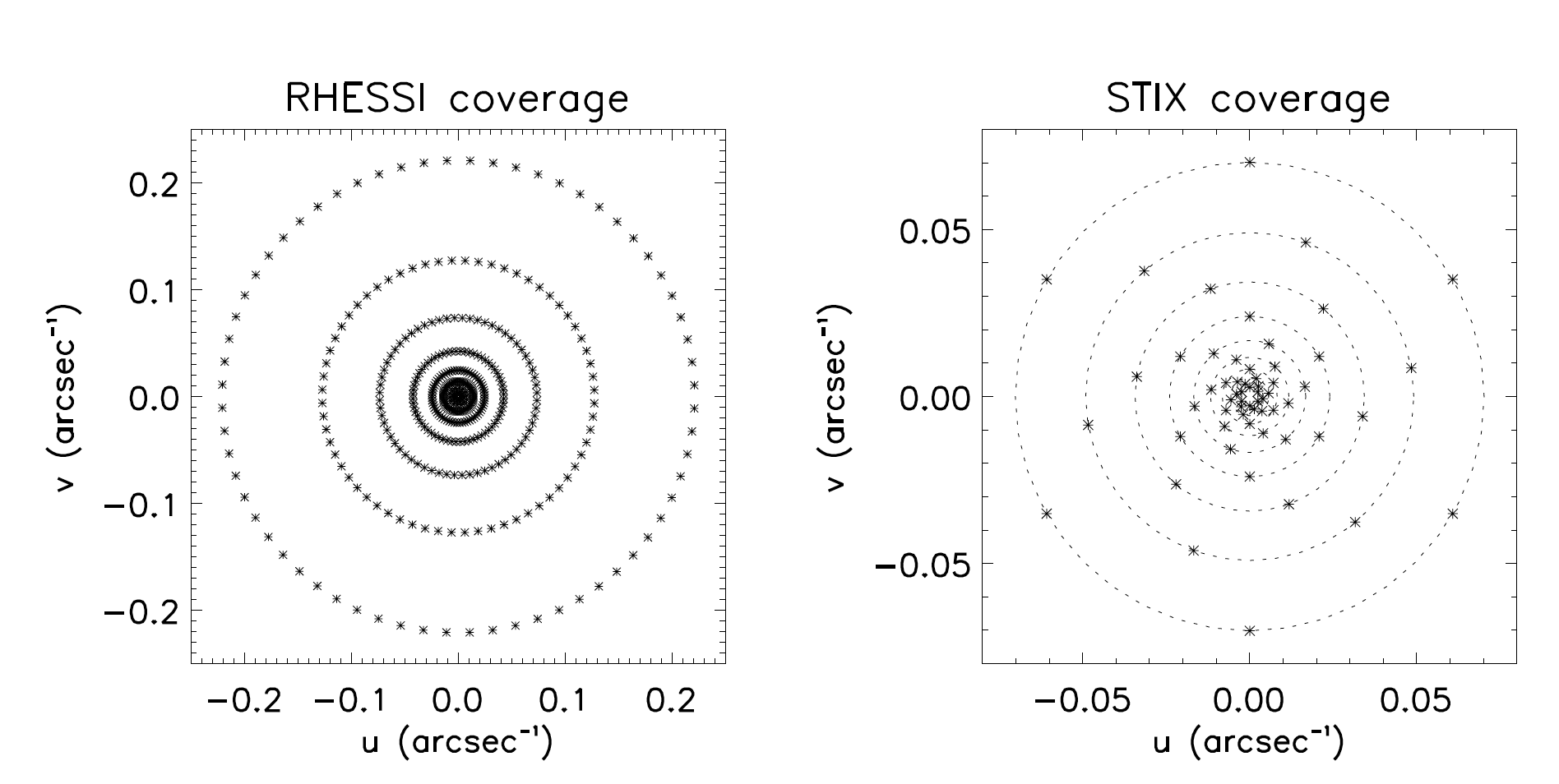} 
\caption{Sampling of the visibility $(u,v)$ plane performed by {\em{RHESSI}} (left panel) and {\em{STIX}} (right panel), respectively.}\label{fig:RHESSI-STIX}
\end{center}
\end{figure}

\begin{figure}[h]
\begin{center}
\includegraphics[width=9.cm]{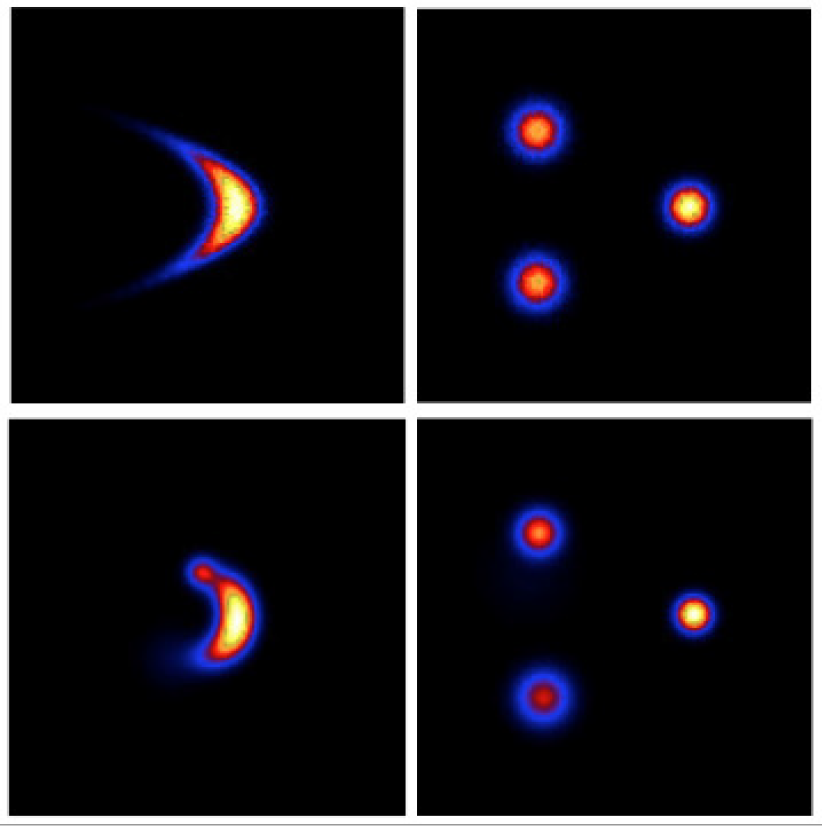} 
\caption{Image reconstructions from {\em{RHESSI}} visibilities provided, from left to right, by a Back-Projection algorithm, CLEAN, an interpolation/extrapolation method, a compressed sensing method based on exploiting an image catalogue, and by our wavelet-based compressed sensing method. The reconstructions refer to the May 13, 2013 event in the time interval 02:04:16-02:04:48 UT and energy range 6-12 keV. {\em{RHESSI}} visibilities recorded by detectors from 3 to 9 have been used in all cases.}\label{fig:VIS-WV}
\end{center}
\end{figure}

Figure \ref{fig:RHESSI-STIX} shows how {\em{RHESSI}} and {\em{STIX}} grids sample the $(u,v)$ frequency domain. From this design, it follows that the mathematical model for data formation in the framework of these two instruments is, in a matrix form:
\begin{equation}\label{a}
HFx = V~,
\end{equation}
where $x$ is the photon flux image to reconstruct, $V$ are the experimental visibilities, $F$ is the discretized Fourier transform, $H$ is a mask that realizes the sampling in the $(u,v)$ plane. The reconstruction of $x$ from $V$ is an ill-posed problem and therefore regularization is required to mitigate the numerical instabilities induced by the observation noise. A possible approach is to apply an $l_1$ penalty term in some transformation domain. This can be realized by solving the minimum problem \cite{ref:duetal18}
\begin{equation}\label{b}
{\hat{x}} = \min_x \{\|HFx - V \|^2_2 + \lambda \|Wx\|_1 \}~,
\end{equation}
where the regularization term $\|Wx\|_1$ is designed to penalize reconstructions that would not exhibit the sparsity property with respect to the Finite Isotropic Wavelet Transform \cite{ref:duetal18}. Figure \ref{fig:VIS-WV} compares the reconstructions provided by this compressed sensing algorithm to the ones obtained by using other four visibility-based imaging methods currently implemented in the {\em{RHESSI}} pipeline \cite{ref:depe09} \cite{ref:feetal17} \cite{ref:maetal09}.

\section{Sequential Monte Carlo for hard X-ray image reconstruction}
\begin{figure}
\begin{center}
\includegraphics[width=6.4cm]{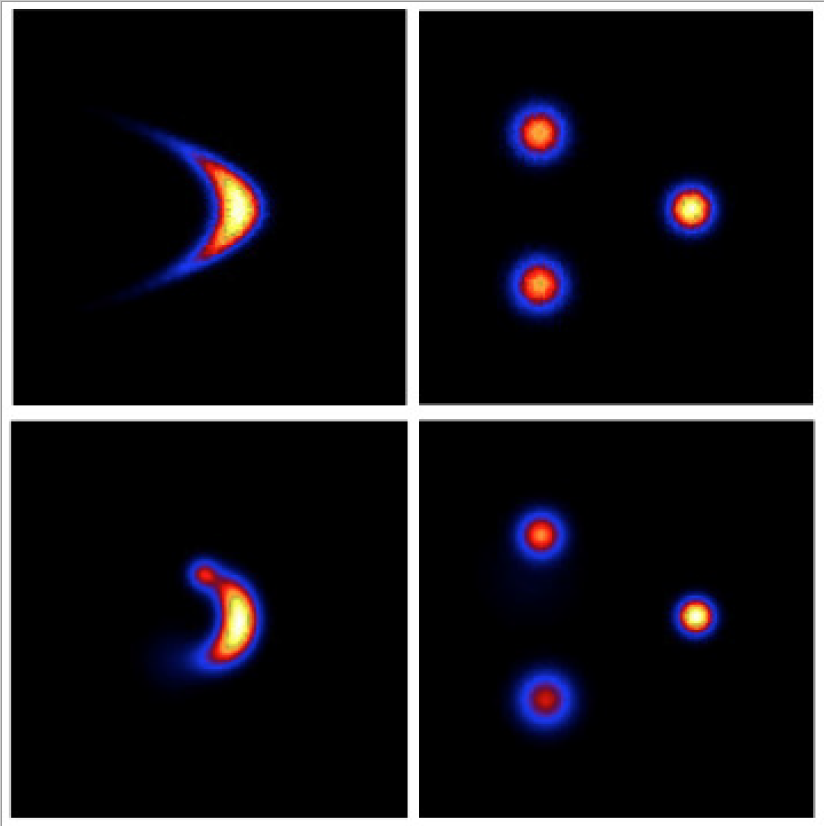}  \\   
\caption{Reconstructions of two simulated configurations estimated by the SMC algorithm using synthetic {\em{STIX}} visibilities corresponding to a realistic signal-to-noise ratio. Top row: ground truth; bottom row: SMC reconstructions.}\label{fig:SMC-simulated}
\end{center}
\end{figure}

\begin{figure}
\begin{center}
\includegraphics[width=11.25cm]{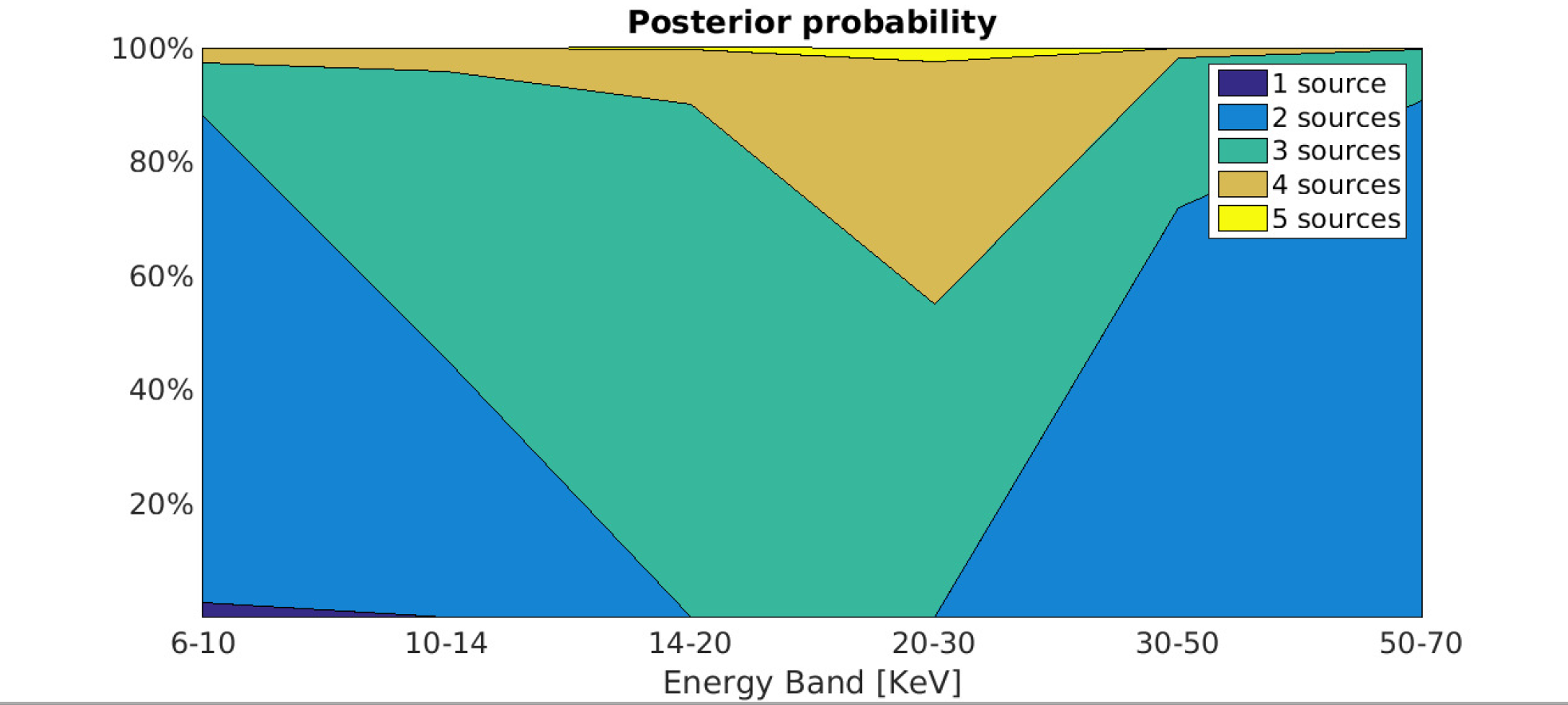}

\hspace{0.5cm}\includegraphics[width=3.9cm]{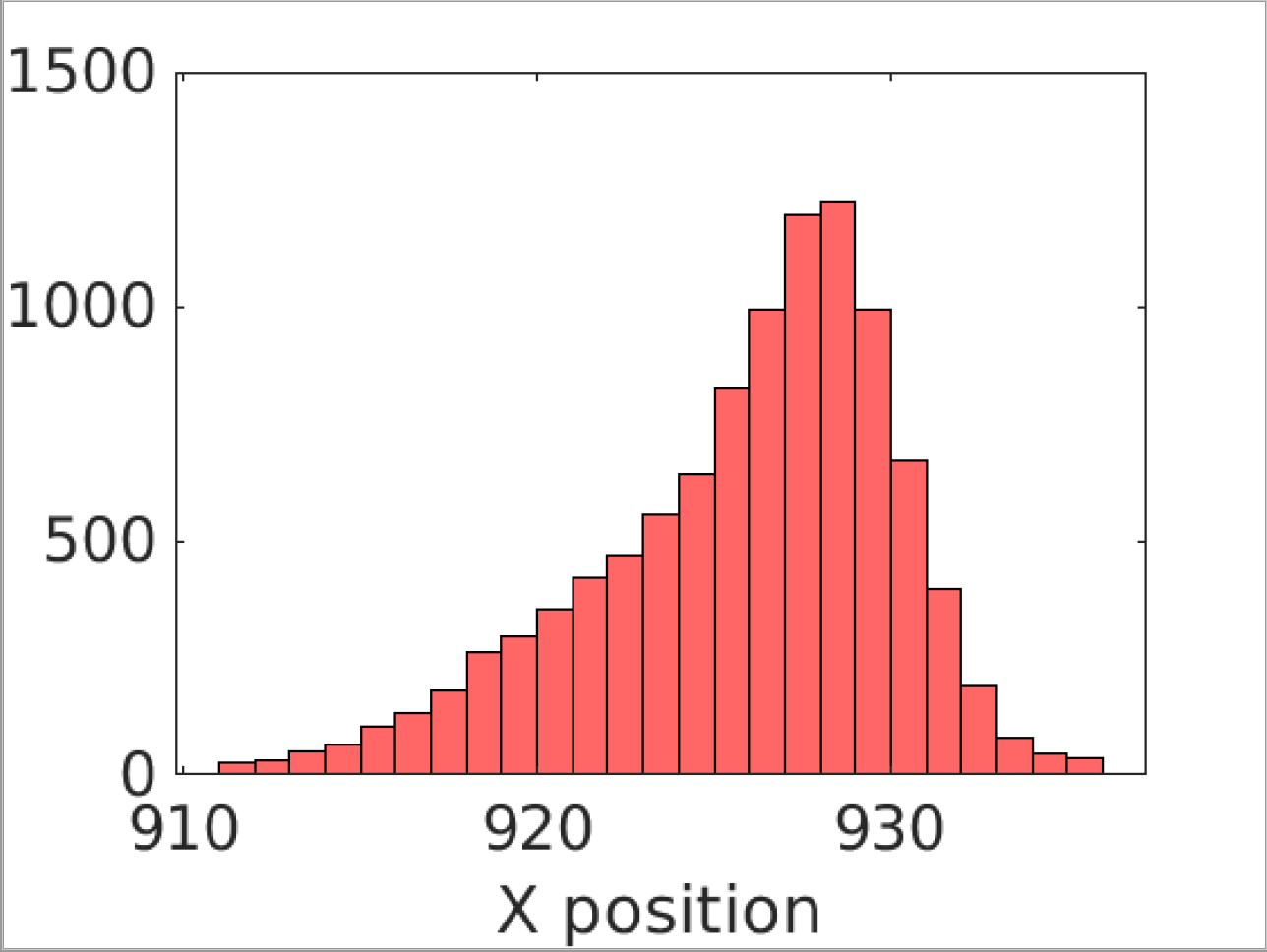} 
\includegraphics[width=3.9cm]{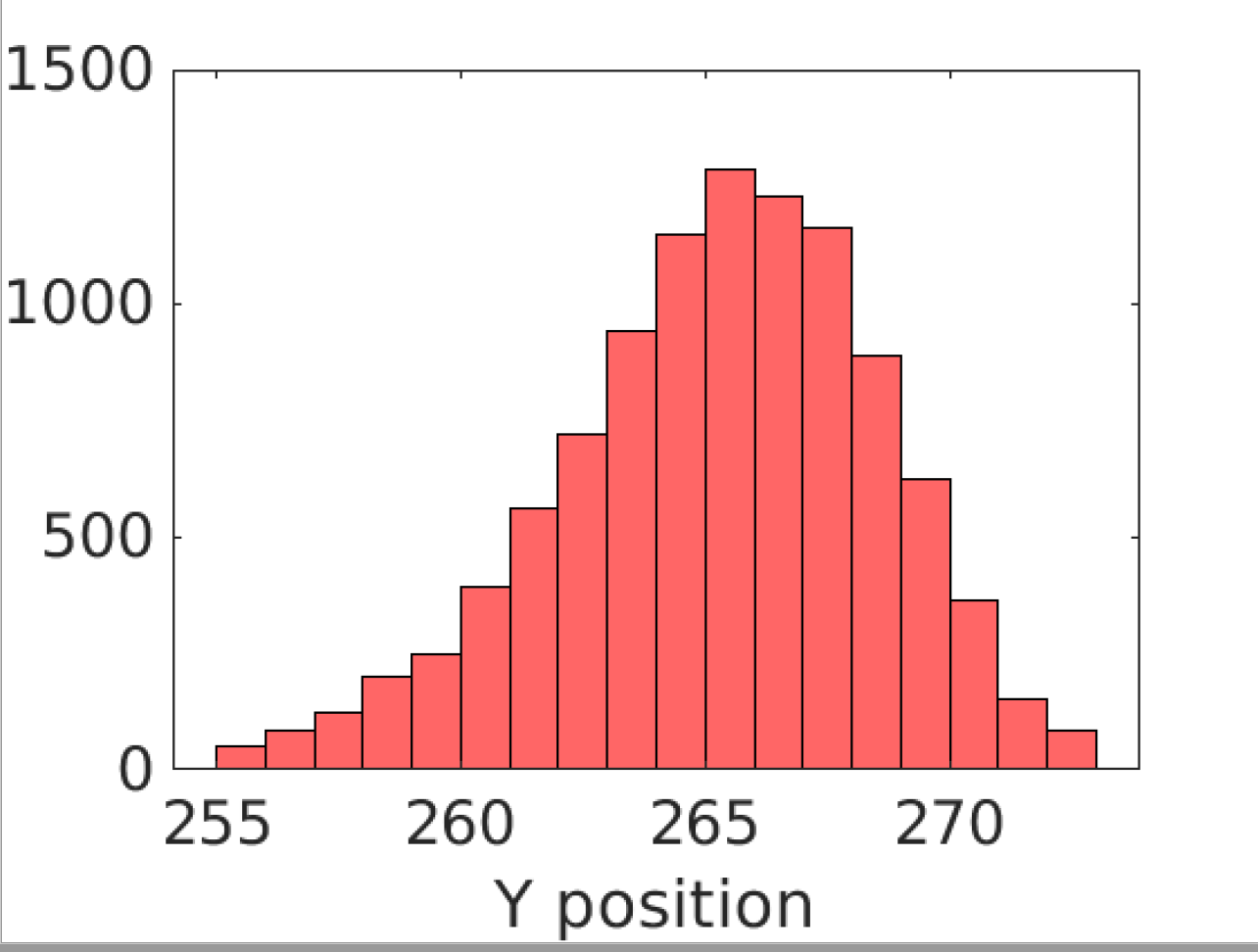}   
\caption{Parameters estimation provided by SMC in the case of the February 20, 2002 {\em{RHESSI}} visibilities. Top panel: posterior probabilities for the number of sources at different energy channels. Bottom panel: histograms of the $x$ and $y$ position of the loop-top source detected with high probability at the $20-30$ keV channel.}\label{fig:SMC-real}
\end{center}
\end{figure}

Sequential Monte Carlo (SMC) samplers are computational methods aiming at sampling target distributions of interest, and are often applied to sample the posterior distribution $p(x|y)$ as given by Bayes' theorem
\begin{equation}\label{c}
p(x|y) = \frac{p(y|x) p(x)}{p(y)}~,
\end{equation}
where $x$ is the unknown, $y$ is the observation, $p(x)$ is the prior probability encoding all {\em{a priori}} information, $p(y|x)$ is the likelihood encoding the image formation model (\ref{a}) and the noise model, and the marginal likelihood $p(y)$ is a normalization factor. In the case of {\em{RHESSI}} and {\em{STIX}} imaging, $x$ is the image to reconstruct and $y$ denotes the set of recorded visibilities. We modeled $x$ as $x(N,T_{1:N},\Theta_{1:N})$ where $N$ is the number of sources in the image, $T_{1:N}=(T_1,\ldots,N)$ represents the source types (Gaussian, elliptical, loop-like) and $\Theta_{1:N}=(\theta_1,\ldots,\theta_N)$ contains the parameters characterizing each source. We chose a prior distribution factorized as the product of a Poisson distribution for $N$, uniform distributions for the source types and uniform distributions for the source parameters \cite{ref:scetal19} \cite{ref:scetal18}. Sequential Monte Carlo \cite{ref:deetal06} computes the posterior distribution iteratively, by constructing a sequence of converging approximate distributions. Once the posterior is determined, it can be used to compute the solution image and all image parameters. Figures \ref{fig:SMC-simulated} and \ref{fig:SMC-real} show results provided by this approach using simulated {\em{STIX}} visibilities and experimental {\em{RHESSI}} visibilities, respectively.

\section{Conclusions}
This paper shows the performances of two image reconstruction methods formulated for hard X-ray solar visibilities. The implementation of the corresponding tools within {\em{Solar SoftWare (SSW)}} is under construction.

\end{document}